\documentclass[preprint,review,12pt,compress]{elsarticle}

\usepackage{blindtext}
\usepackage{geometry}
\usepackage{amssymb} 
\usepackage{lipsum}
\usepackage{xcolor}
\usepackage{verbatim}
\usepackage{float}
\usepackage{soul}
\usepackage[final]{hyperref} 
\hypersetup{
	colorlinks=true,       
	linkcolor=blue,        
	citecolor=blue,        
	filecolor=magenta,     
	urlcolor=blue
}
\usepackage{lineno}

\journal{Nuclear Instruments and Methods in Physics Research B}
\begin{document}

\begin{frontmatter}

\title{
The IAEA electronic stopping power database: modernization, review, and analysis of the existing experimental data}

\author[label1]{C. C. Montanari}
\author[label2]{P. Dimitriou}
\author[label2]{L. Marian}
\author[label1]{A. M. P. Mendez}
\author[label1]{ J. P. Peralta}
\author[label1]{F. Bivort-Haiek}

\address[label0]{Instituto de Astronom\'{\i}a y F\'{\i}sica del
Espacio, CONICET -- Universidad de Buenos Aires, \\ Buenos Aires, Argentina.

\affiliation[label1]{
             organization={Instituto de Astronomía y Física del Espacio. CONICET and Universidad de Buenos Aires},
             addressline={IAFE, Ciudad Universitaria},
             postcode={1428},
             city={Buenos Aires},
             country={Argentina}
             }
}

\affiliation[label2]{
             organization={Division of Physical and Chemical Sciences, International Atomic Energy Agency},
             addressline={Vienna International Centre, P.O. Box 100},
             postcode={A-1400},
             city={Vienna},
             country={Austria}
             }

\begin{abstract}
We review the electronic stopping power data within the IAEA database, assessing the abundance and scarcity of available data as a function of energy and collisional systems. Our analysis includes an examination of 
the experimental values, their evolution in time, the dispersion among data, and trends of the more recent measurements. Additionally, we provide comparisons with SRIM-2013 calculations for select cases of interest. Notably, we identify sparsely measured systems and energy regions, emphasizing the pressing need for new, reliable data and independent theoretical predictions.
\end{abstract}


\end{frontmatter}


\section{Introduction}
\label{introduction}

Electronic stopping power data are relevant to a wide range of applications, from ion beam analysis, deposition ranges, ion implantation, and radiation damage to medical studies and treatments. Reliable stopping powers are also needed in isotope production for medical applications, fusion technologies, non-destructive assay for nonproliferation, and detector developments. By the end of 2015, the International Atomic Energy Agency (IAEA) became responsible for hosting, maintaining, updating, and improving the electronic stopping power database \cite{iaea,MONTANARI201750}. This database is the legacy of Helmut Paul \cite{PAUL2003,PAUL2009,PAUL2013_AIP}, a pioneer of this field, who, in the 1990s, had the vision of making all stopping power data available to the whole scientific community. 
The database includes almost a century of
experimental measurements in different laboratories worldwide, including the early measurements by Rosenblum in 1928 \cite{Rosenblum28} for alpha particles in Li, Cs, Zn, Pd, and Pb, and by Bätzner in 1936 \cite{Batzner1936} for hydrogen in Al, Cu, Sn, Ag, and Au. Since 2015, the database has been periodically updated with newly published data~\cite{iaea}.  

The IAEA Stopping Power database currently contains experimental values for 1526 collisional systems, of which around 60\% (913) have only one dataset,  while 233 have two datasets, which do not necessarily overlap in energies. These statistics indicate
that less than $30\%$ of the systems in the database can be considered reasonably well-known from an experimental point of view.  
Most of the compiled experimental measurements have been performed on solids, but there are also data for gases and liquids, single atomic targets, and compounds.

A large number of systems with a single dataset is well described by semi-empirical methods, such as SRIM-2013~\cite{srim}. This agreement is expected but does not represent an actual independent comparison. 
SRIM-2013 calculations agree with data published before 2013 within 5-10\% for all ions~\cite{srim_article,ziegler_book}. The agreement worsens to an average of 11.4\% (19.2\% for H ion, 10.6\% for He ion, and 6.6\% for the rest) for measurements performed after 2013, which were not considered in the empirical fit~\cite{ESPNN_2022}. 
Despite the mentioned inaccuracies, 
SRIM-2013 is used in the analysis of experimental data, for example, to determine the target thickness and inhomogeneities, to evaluate contaminant contributions, to normalize data at very high energies, or in Monte Carlo simulations to separate nuclear and electronic stopping. A comprehensive analysis of SRIM, including weaknesses and misconceptions, can be found in Wittmaack~\cite{WITTMAACK2016}.

The stopping of slow ions in matter is a major topic of research in fundamental and applied physics. 
Nonetheless, there are still scarce measurements in this velocity region, mainly for heavy ions. 
The low-velocity stopping depends on the electronic configuration and type (metal or insulator, solid or gas) of target. For compounds, the Bragg rule~\cite{bragg1905,bragg} is usually employed; however, in the low ion-velocity region, this empirical rule is unreliable.

The experimental determination of the electronic stopping power of slow ions is challenging due to the sensitivity to target characterization (contaminants, inhomogeneity in thickness, crystallinity)~\cite{BRUCKNER2021}, the complex dynamics involved, the dependence on the impact parameter, and the difficulty in separating the nuclear component from the electronic one.  From the theory perspective, determining the energy loss of slow-heavy ions in solids requires implementing highly non-perturbative approaches and impact-parameter considerations~\cite{Sigmund2023,Sigmund2021}; in some cases, one may also have to consider the probability of energy loss in autoionization states due to molecular orbital promotion~\cite{ZINOVIEV2024}. As demonstrated in this work, these complexities lead to sparse and scattered values.

This contribution aims to review the available electronic stopping power data by considering the abundance or scarcity of data, preferred ion-target systems in recent measurements, trends, and data reliability. We highlight the need for new measurements (ion-target systems, energy regions) and models with descriptive and predictive power
but also recognize the importance of assessing the existing data and making recommendations to the user community.

In Section~\ref{sec:database}, we present the evolution of measured systems and data acquisition over time. 
In Section \ref{sec:examples}, we focus on specific examples of interest. Notably, we examine water, given its relevance in medical applications, and explore oxides, which have seen numerous recent measurements, i.e., Ta$_2$O$_5$ and TiO$_2$. Challenges, limitations, differences among data sets, and energy regions without values are discussed in Section~\ref{sec:chall}. Finally, we summarize our conclusions in Section~\ref{sec:concl}.

\section{Data dissemination and exploration}
\label{sec:database}

\begin{figure}[t]
\centering 
\includegraphics[width=\textwidth]{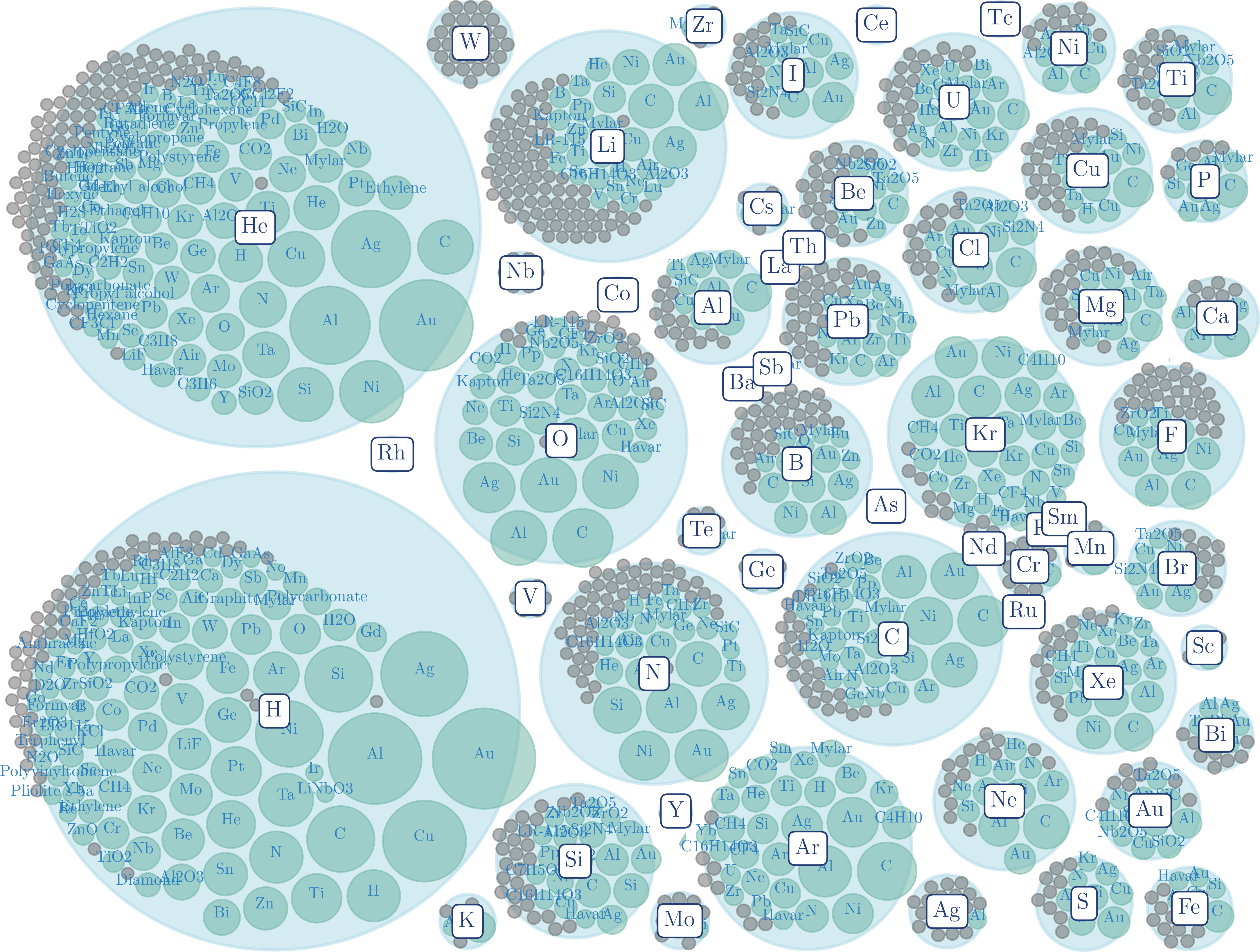}

\vspace{-0.1cm}
\caption{
Circular packing of stopping power data available in the database; the size is proportional to the number of datasets. Light blue circles correspond to projectiles (ion bubbles). Blue circles inside each ion bubble represent values for different targets. Grey bubbles correspond to targets with a single dataset.}
\label{fig:dataset-bubbles}%
\end{figure}

The IAEA Stopping Power database~\cite{iaea} was recently modernized. 
The data content was kept practically the same, although several improvements were made to complete relevant descriptors.
The data files and metadata were sorted, restructured, and coupled to an API, leading to a more versatile and dynamic exploration of the content of the database (including data and references) 

and flexible retrieval of the data. 
 
The layout of the webpage is now more user-friend; allll the systems can be filtered 
on-demand by ion-target combination or author. 
The data is displayed through interactive plots and comprehensive data tables that include information on the target phase, uncertainties, units, and references. The data tables can be downloaded as text files or CSV files. Hyperlinks to online publications are also included. Users can easily access the data and publications through the 
interface or directly via the API for seamless retrieval.

Figure \ref{fig:dataset-bubbles} shows a circular packing plot of the existing experimental stopping power data. The light blue circles, called ion bubbles, represent the data for a given ion; the dimension of the circles is proportional to the number of datasets of said ions as projectiles.
It is evident from this figure that most of the measurements correspond to H and He ions, followed by O, Li, and N. The distribution of targets per ion is illustrated in the figure with blue circles inside each ion bubble. The size of each target bubble is also proportional to the number of datasets. Evidently, H and He ions on Al, Au, and Ag targets are among the collisional systems with the most measurements, with over 40 and up to 70 independent datasets.  We found that the $58\%$ of the systems are composed of atomic targets while $42\%$ are compounds.
We draw attention to these statistics and stress the importance of measuring stopping in compounds for expanding the knowledge beyond the limitations of the Bragg additive rule~\cite{bragg1905}. 
In Figure~\ref{fig:dataset-bubbles}, we highlight with dark grey the single dataset per ion-target combinations. As mentioned before, the collisional systems with 
only one set of measurements amount to 60\%; however, we can see from the figure that there are specific ions, such as W, Mo, and Ag, for which the large majority of measurements are single datasets.
It is worth noting that certain atomic elements, including Na, P, K, S, I, Ba, Os, As, Pm, Eu, and Tl, lack experimental stopping values for any impinging ion. A thorough graphical representation of the database is depicted with a matrix plot in Fig.~1 of the supplementary material attached to this work. This figure allows the reader to visualize many of the features mentioned above, but most importantly, it shows the large gaps in the stopping power experimental data.

\begin{figure}[t]
\centering 
\includegraphics[width=\textwidth]{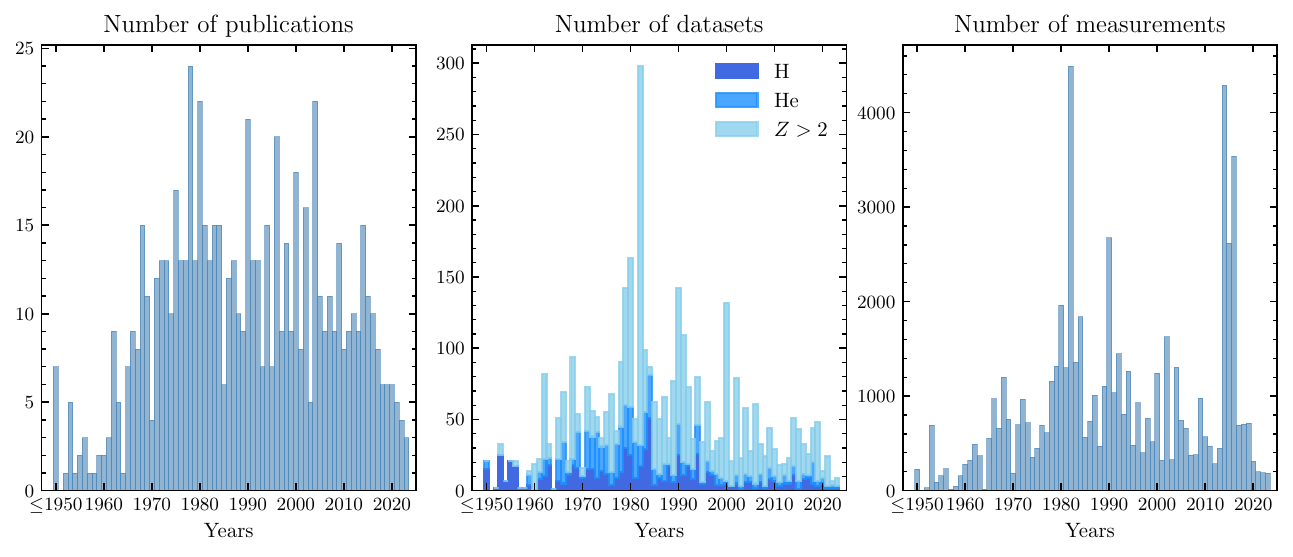}

\vspace{-0.4cm}
\caption{
Distribution as a function of time of publications including experimental electronic stopping power (left), number of datasets measured (center), and number of data point measurements (right).}
\label{fig:dataset-years-ions}%
\end{figure}

A representation of the evolution of the experimental data as a function of time is shown in Figure~\ref{fig:dataset-years-ions}, where we consider the publication date, the number of datasets measured, and the number of measurements, i.e., data points. 
The left plot in Figure~\ref{fig:dataset-years-ions} displays the number of publications measuring absolute values of the stopping power as a function of ion energy.
Publications on other related topics, such as relative measurements, differential stopping cross-sections, simulations, extrapolations, applications, and theoretical models, are not considered. 
The center plot shown in Figure~\ref{fig:dataset-years-ions} distinguishes the number of datasets by ion: H, He, and heavier ones. It is noted that most of the measurements since 1980 correspond to heavy ions.  
The number of measurements, or data points, displayed in the right plot of Figure~\ref{fig:dataset-years-ions} reveals marked peaks in 1982 and 2014-2016. The first peak corresponds to the large number of new datasets published in 1982 (as shown in the middle plot). 
The subsequent peak represents an increase in the number of measurements per dataset, which may be related to advancements in experimental techniques and the systematic approach to data acquisition~\cite{FONTANA2016}.

As a general comment on Figure~\ref{fig:dataset-years-ions}, we notice that despite the lack of stopping data already mentioned and the partial energy range coverage, the number of publications, measured datasets, and data points has decreased since 2015.
Likely, the overconfidence in the stopping power values provided by SRIM-2013~\cite{srim,srim_article,ziegler_book} could explain this phenomenon. 
It is a fact that SRIM-2013 code~\cite{srim} offers stopping power values for any ion-target system, even for compounds not measured yet by means of the Bragg rule~\cite{bragg}. 

\section{Examples of interest}
\label{sec:examples}

In this section, we present an in-depth analysis of three compounds of interest: water, titanium oxide, and tantalum oxide. 
Throughout this work, the experimental values in all figures adhere to the IAEA stopping power database nomenclature for labeling. This nomenclature combines the authors' surnames with the publication year.
For example, the stopping power measurements of protons in liquid water by Siiskonen \textit{et al.}~\cite{Sn2011}, published in 2011, are labeled as Sn2011. All the references displayed in the figures can be found in the IAEA Stopping Power database~\cite{iaea}.

\subsection{Stopping power of water}

\begin{figure}[t]
\centering 
\includegraphics[width=0.55\textwidth]{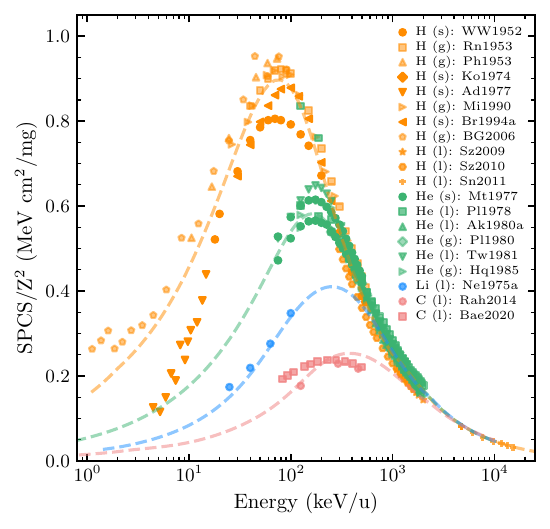}

\vspace{-0.6cm}
\caption{(color online) Scaled stopping cross-section of H, He, Li, and C ions on H$_2$O. 
Symbols: experimental data of target in solid (s), liquid (l), and gaseous (g) states~\cite{iaea}. 
The labels adhere to the IAEA database nomenclature, which is a combination of references to the authors' surnames and the year of publication. Curves: semi-empirical SRIM-2013 values~\cite{srim}.}
\label{fig:water}%
\end{figure}

The case of water is interesting for biological targets and medical applications, mainly hadron-therapy~\cite{cancer2022,Geant4_stopping}. Only four ions in water feature energy loss experimental data are available: H, He, Li, and C, as shown in Figure~\ref{fig:water}. 
The stopping cross-section (SCS) has been scaled with $Z^2$, which provides an interesting overlapping of the different ion data at high energies (per nucleon). In this figure, we include the experimental data for gas (g), solid (s), and liquid (l) water with markers of different colors per ion and type per dataset. 

The dashed curves in Figure~\ref{fig:water} correspond to SRIM-2013~\cite{srim} for gas water. 
The measurements of protons in water cover an extended energy range and different water phases. However, the other cases are sparsely measured. Carbon in water is the system less examined, which is surprising considering its importance in carbon ion therapy~\cite{cancers9060066, hadrontherapy2020, hadrontherapy2015}. 
From Figure~\ref{fig:water}, one can make the following remarks: solid-gas-liquid differences are significantly below the stopping maximum, while they are negligible for energies above 1~MeV/amu. In the low energy region, the measurements for solid water date from the 50s and 70s~\cite{MONTANARI201750}. New measurements for protons in ice with up-to-date techniques would be very interesting. The scaled stopping cross-sections allow us to inter-compare different ions and phases, which is an extra test for the experimental data. The data for He and Li ions is forty years old and has already been considered in the semi-empirical SRIM-2013 fit. On the other hand, the recent data for C ions for liquid water~\cite{Baek2020} suggests a different stopping maximum compared to SRIM-2013. The importance of C makes this ion worthy of being examined more extensively from an experimental perspective at higher and lower energies. Independent theoretical calculations should accompany these studies, and ideally, all the available data should be critically assessed to reveal possible deficiencies in the measurements and lead to recommended experimental values.

\subsection{Stopping power of oxides}

\begin{figure}[t]
\centering 
\includegraphics[width=0.55\textwidth]{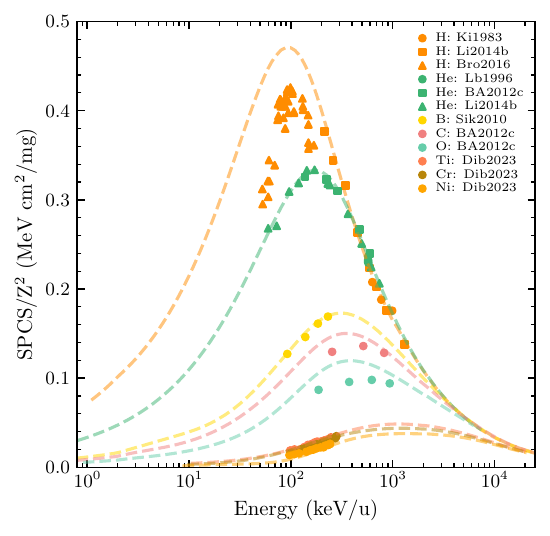}

\vspace{-0.6cm}
\caption{(color online) Scaled stopping cross-section 
of H, He, B, C, O, Ti, Cr, and Ni ions on TiO$_2$. 
Symbols, curves, and labels are the same as in Fig.~\ref{fig:water}.
}
\label{fig:tio2}%
\end{figure}
\begin{figure}[h!]
\centering 
\includegraphics[width=0.55\textwidth]{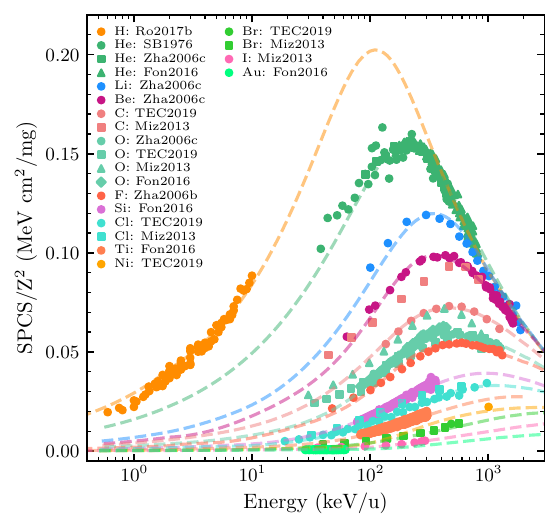}

\vspace{-0.6cm}
\caption{(color online) Scaled stopping cross-section 
of H, He, Li, Be, C, O, F, Si, Cl, Ti, Ni, Br, I, and Au ions on Ta$_2$O$_5$. 
Symbols, curves, and labels are the same as in Fig.~\ref{fig:water}.
}
\label{fig:ta2o5}%
\end{figure}

The stopping in solid oxides has received both experimental and theoretical attention, with numerous measurements performed for Al$_2$O$_3$ (17 different ions), SiO$_2$ (14 ions), and Ta$_2$O$_5$ (14 ions), but also for ZrO$_2$ (9 ions), Nb$_2$O$_5$ (8 ions), TiO$_2$ (5 ions), HfO$_2$ (5 ions), Er$_2$O$_3$ (H and He ions), VO$_2$ (H ions), WO$_3$, UO$_2$, and SO$_2$ (only measurements for He ions), and SnO$_2$ (only for F ions). 
The data for different ions in aluminum oxide are shown in Figure~4 of \cite{MONTANARI201750}. In the present work, the scaled stopping cross-section of all the available measurements for TiO$_2$ and Ta$_2$O$_5$ are displayed in Figures \ref{fig:tio2} and \ref{fig:ta2o5}, respectively. We showcase these oxides due to a recent increase in measurements, as documented in the literature~\cite{DIB2023, TEC2019, Bro2017, FONTANA2016, Limandri2014b, Miz2013}. The figures illustrate uneven energy-range coverage in these measurements.
In Figure~\ref{fig:tio2}, we note the differences between the experimental data and SRIM-2013 for H ions \cite{Bro2017} (Bro2017), C and O ions \cite{BA2012c} (BA2012c),
and Ni ions \cite{DIB2023} (Dib2023). Remarkably, the new data for Ti and Cr ions in TiO$_2$ from the very recent work of Dib and co-workers~\cite{DIB2023} agree very well with the SRIM-2013 predictions, even though the latter did not consider these recent measurements as it predates the measurements by 10 years.

The 14 ions measured in Ta$_2$O$_5$ are included
in the single plot in Fig.~\ref{fig:ta2o5}.  The large amount of data makes it difficult to see any details; however, the low-energy regions with and without data are very clear. Notably, the data of Roth et al.~\cite{Roth2017} (Roth2017b) agree with SRIM-2013 predictions, considering that the latter uses the Bragg rule in the low-energy region. SRIM-2013 agrees with the data around the stopping maximum, except for C and O ions measured by Mizohata et al.~\cite{Miz2013} (Miz2013).

\section{Challenges}
\label{sec:chall}

There are many open challenges in the field of stopping power. We draw attention to some of the issues that, to our understanding, require attention: The first one is
the lack of data for many systems. The matrix plot of the supplementary material shows striking blanks in the measured systems. For ions heavier than Li, the molecular targets are considerably under-measured. The figure also illustrates examples of compounds of technological interest that deserve more experimental and theoretical efforts, particularly nitrides, carbides, and borides. 
These groups (N, C, and B combined with metals) are used in many applications, such as solar cells, coating materials, cutting tools, and protective layers in nuclear reactors. The stopping power measurements in carbides and nitrides drew attention in the last twenty years (from \cite{zha2003b} in 2003 to \cite{Sortica2019} in 2019) due to their extraordinary properties, which compare 
to pure metals~\cite{NINGTHOUJAM2017337}; however, the database only includes values for SiC, GaN, InN, TiN, and Si$_3$N$_4$, with SiC being the most measured and the only carbide with stopping data.

Another topic that requires attention is
the extended use of the Bragg rule to describe molecules by combining the data of its constituent atoms~\cite{bragg1905,bragg}, even at low impact energies. This approximation is combined with
the use of stopping values of sparsely-measured (or even unmeasured) atomic systems to predict stopping in molecules, resulting in the combination of two possible factors of uncertainty.

In what follows, we particularly highlight and analyze three open subjects based on the available values in the database: (1) the large number of single-dataset systems 
and the comparison with semiempirical SRIM-2013; (2) the important differences between new measurements, previous data, and SRIM-2013 predictions around the stopping maximum, which calls for a critical assessment of the data; and (3)
the small fraction of systems with stopping power measurements at low energies.

\subsection{Single dataset systems and SRIM}

As previously mentioned, 60\% of measured systems have only one dataset. The absence of independent confirmation of these measurements highlights our limited knowledge of stopping power in a large number of ion-targets.

\begin{figure}[t]
\centering 
\includegraphics[width=\textwidth]{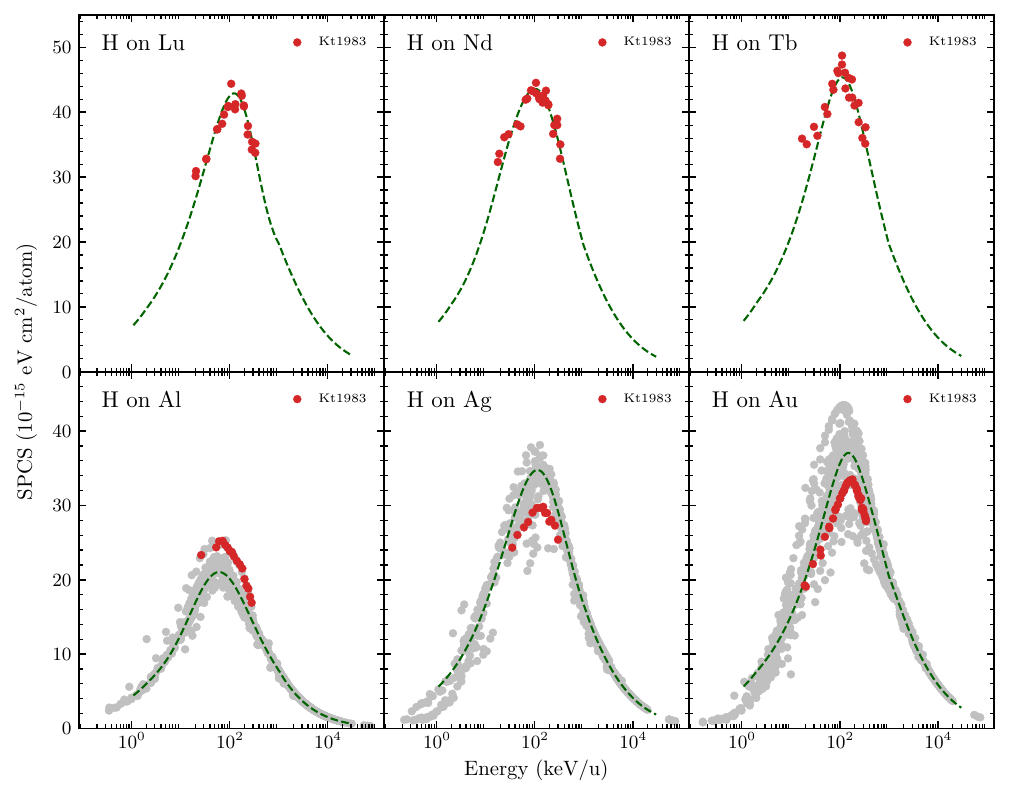}

\vspace{-0.6cm}
\caption{Stopping cross-section
of H on Lu, Nd, Tb, Al, Ag, and Au. Experimental data \cite{iaea} (symbols) and semi-empirical SRIM-2013 values \cite{srim} (dashed curves). Measurements by Krist and Mertens from 1983~\cite{Kt83a} are illustrated in red. }
\label{fig:Krist}%
\end{figure}

In the top row of Figure~\ref{fig:Krist}, we illustrate this situation with subplots for three such cases: protons in lanthanides Lu, Nd, and Tb. The measurements by Krist and Mertens from 1983~\cite{Kt83a} (Kt1983) are displayed with red bullets. We also included the SRIM-2013 results with dashed curves. The agreement of SRIM-2013 with the experimental data is almost perfect. 
However, this publication \cite{Kt83a} also provides measurements of the same ion on many other targets. Three of them are the most measured collisional systems in the history of stopping power: protons in Au, Ag, and Al. The bottom row of Figure~\ref{fig:Krist} presents with red bullets the results in Ref.~\cite{Kt83a} for protons in Al, Ag, and Au. This data disagrees with most of the experimental values for the three targets. Moreover, the available experimental stopping of Lu, Nd, and Tb cannot be described thus far by theoretical models such as CASP \cite{casp6.0,CasParticle}, DPASS \cite{dpass,dpass_article}, and SLPA \cite{Peralta2023}. These circumstances make it imperative to carry out new measurements in lanthanides and perhaps in many more single-set systems.

\subsection{New data versus the \textit{status quo}}

\begin{figure}
\centering 
\includegraphics[width=\textwidth]{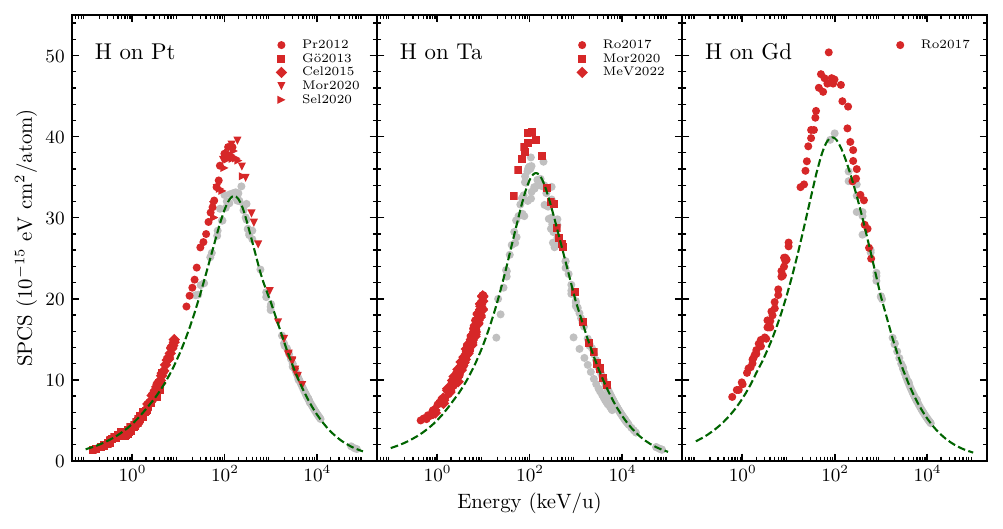}

\vspace{-0.6cm}
\caption{Stopping cross-section 
for H on Pt, Ta, and Gd. Experimental data \cite{iaea} (symbols) and semi-empirical SRIM-2013 values \cite{srim} (dashed curves). Recent measurements \cite{Pr12,Go13,Ro17,Mor20,Cel15,Sel20,MeV22} are illustrated in red.}
\label{fig:newdata}%
\end{figure}

In some cases, recent stopping power measurements differ from the present status of knowledge represented by the previous measurements and SRIM-2013. 
In Figure~\ref{fig:newdata}, we display the cases of protons in Pt, Ta, and Gd. 
The data published from 1967 to 1996 is shown in grey symbols, while we illustrate in red the measurements from 2012 to date by the groups led by Primetzhofer (Uppsala)~\cite{Pr12,Go13,Ro17,Mor20}, Arista (Bariloche)~\cite{Cel15}, Grande (Porto Alegre)~\cite{Sel20}, and Valdés (Valparaíso)~\cite{MeV22}. 
The disagreement around the maximum with the older data and the discrepancies with the SRIM-2013 curves are clear. These differences are reasonable since SRIM-2013 is a model with fitted parameters that efficiently describe the data published up to its latest version, which dates over a decade ago. 
On the other hand, new full theoretical works describe quite well these recent measurements~\cite{Peralta2023,Sel20,Li2022,deVera2023,Peralta2022}. It is clear that in these cases, both the old and new data need to be critically assessed, outliers identified and recommended values offered to the model developers and end users.

\subsection{Stopping power for slow ions}

\begin{figure}[t]
\centering
\includegraphics[width=\textwidth]{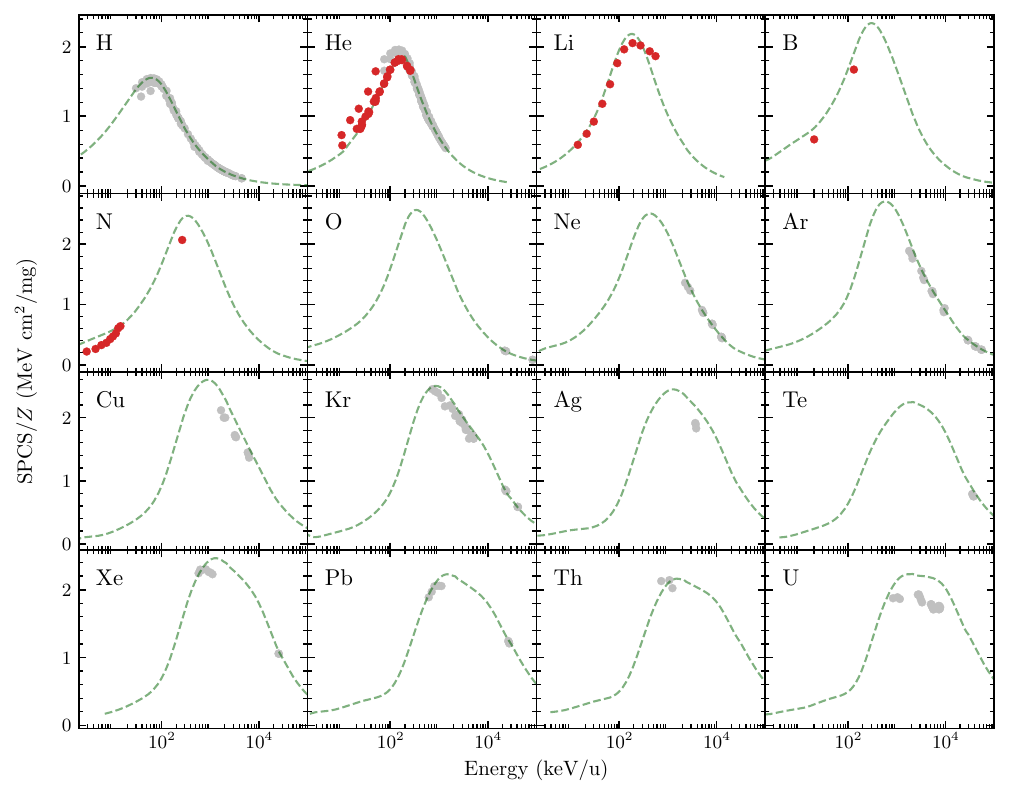}

\vspace{-0.6cm}
\caption{Stopping cross-section of CH$_4$ for 16 ions. Experimental data \cite{iaea} (symbols) and semi-empirical SRIM-2013 values \cite{srim} (dashed curves). Slow ion datasets are illustrated in red (see text).}
\label{fig:slowions_CH4}
\end{figure}

The determination of the electronic stopping power for low-velocity projectiles is complex, i.e., one needs to consider a many-body description, non-perturbative approaches, trajectory dependence, and sensitivity to target characterization. Moreover, the nuclear stopping is not negligible and must be subtracted from the total measured values. Recent theoretical works cast doubts on the separation of nuclear and electronic stopping based on the Coulomb deflection of the ion at very low energies and the impact-parameter dependence of the electronic stopping \cite{Sigmund2021,SIGMUND2017}, or discuss the influence of molecular orbital promotion in the energy loss of slow heavy-ions in matter~\cite{ZINOVIEV2024}. 
These contributions may lead to a misguided estimation of the inelastic part of the total stopping,  deviations from velocity-proportionality, and isotope effects.

At present, the available data for low-energy stopping (per nucleon)  is scarce; however, there have been many experimental efforts in the last ten years by different scientific groups and laboratories around the world: Uppsala University (Sweden)~\cite{Roth2017, Ro17, SHAMSLATIFI2023, STROM2021, MORO2021, Kantre21, Lohman2020, Tran2020, KANTRE2019, BRUCKNER2018, Sortica2017, NAQVI2016, Pr2013}, Universidad T\'ecnica Federico Santa Mar\'ia (Chile)~\cite{MeV22, MeV21,Valdes16}, iThemba LABS~(South Africa) \cite{SIMON2014,GUESMIA2016,DIB2015}, Johannes Kepler University Linz (Austria)~\cite{Go13,ROTH2013,Go14},  University of Jyväskylä (Finland)~\cite{ECHLER2014,ECHLER2017}, Centro Atómico Bariloche (Argentina)~\cite{Cel15,Cel13}, University of Tennessee (USA)~\cite{FONTANA2016,Jin2014},  University of São Paulo (Brazil)~\cite{LINARES2017}, and Colorado School of Mines (USA)~\cite{JEDREJCIC2018}.

Around 52\% of the collisional systems in the database lack stopping power measurements for slow ions. Out of this 52\%, when considering the targets, 47\% correspond to atomic systems and 53\% to compounds; with regards to the projectiles, 25\% are for light ($Z \leq 4$), 58\% for medium ($5\leq Z\leq 54$), and 13\% for heavy ($Z>54$) ions. To obtain these percentages, we analyzed the entire database, and for each collisional system, we found the stopping power maximum (experimental or the one given by SRIM).
Then, we regarded all the datasets with at least one value in the energy region below half the stopping maximum as slow ion measurement. 
In Figure~\ref{fig:slowions_CH4}, we present an example of the present analysis for all the available ions on CH$_4$. The markers colored with red are the ones considered as slow ion values. Note that some sets in red include measurements for low energies but also near the stopping maximum (He, Li, and N), and they have been included together since they belong to the same dataset. In the case of CH$_4$, only 25\% of the systems have measurements below half the maxima. However, other targets --such as Au (45 ions) or Ag (36 ions)-- have around 80\% of their collisional systems featuring slow ion measurements.
These statistics, together with
the discrepancies observed among recent (last ten years) and older stopping data underscore the importance of measuring stopping cross sections for slow ions
and conducting a comprehensive assessment of the techniques and quality of the various data.

\section{Summary and conclusions}
\label{sec:concl}

The modernization of the stopping database created by Helmut Paul in the 1990s makes it possible to mine its rich content efficiently and analyze various aspects of the experimental data, from the large number of data available to the gaps for certain collisional systems. 

In this work, we focused our attention on some cases of interest. The behavior of ions in water and the differences observed in solid, gas, and liquid phases remain open questions, with the case of carbon in water standing out because of its importance for applications. The studies on oxides also show the contrast between the number of measurements and ions versus the not-yet-covered energy regions. The unexpected validity of the Bragg rule around the stopping maximum and below in most cases is noteworthy.

We mentioned, at least briefly, some of the open challenges in the field. We presented the statistics of systems with single datasets and slow velocity ions, which show the limitations and gaps in the available experimental values. 
SRIM-2013's agreement in systems with few datasets is expected. However, by examining similar measurements in other cases with many independent measurements (like Au, Al, or Ag), we cast doubts on some of the single datasets. We consider this not a problem of the SRIM-2013 predictions but of the lack of data. To elucidate these differences, new measurements are required. 
Discrepancies among recent and old measurements around the stopping maximum and below are also noted. They are reasonable because they reflect the results of new techniques and continuous efforts. On the other hand, they mandate a critical assessment of all the techniques and a careful evaluation of the data with uncertainty quantification before the data are used to develop models or implemented directly in applications. 
Although experimental techniques have improved significantly along with our knowledge of the electronic stopping power, the field still faces challenges, particularly in addressing the striking gaps present in the measured systems. 
New and reliable measurements, along with independent theoretical predictions, are encouraged to
further advance the field of electronic stopping power.

\section*{Acknowledgements}
This work was partially supported by the following projects: PIP11220200102421CO by CONICET Argentina, PICT-2020-SERIE A-01931 by ANPCyT Argentina, and the IAEA Coordinated Research Project F41035, Contract No. 28067. FBH acknowledges support from the IAEA in the form of two Special Service Agreements awarded from March to July 2022.

\bibliography{stopping_IAEA}

\end{document}


\begin{center}
\textbf{Supplementary material: Matrix plot of the database}
\end{center}

Fig. \ref{fig:matrixplot} presents a matrix plot of all the experimental data available 
in the IAEA database.
The $x$-axis of the figure shows the projectiles while the $y$-axis presents, 
from row 1 to 92, atomic targets ordered by atomic number and, from row 93 
onward, compound targets are ordered alphabetically. The color intensity of 
each matrix element illustrates the amount of datasets available. 
The actual number of datasets is also given.

\begin{figure}[H]
\centering
\includegraphics[width=0.95\textwidth]{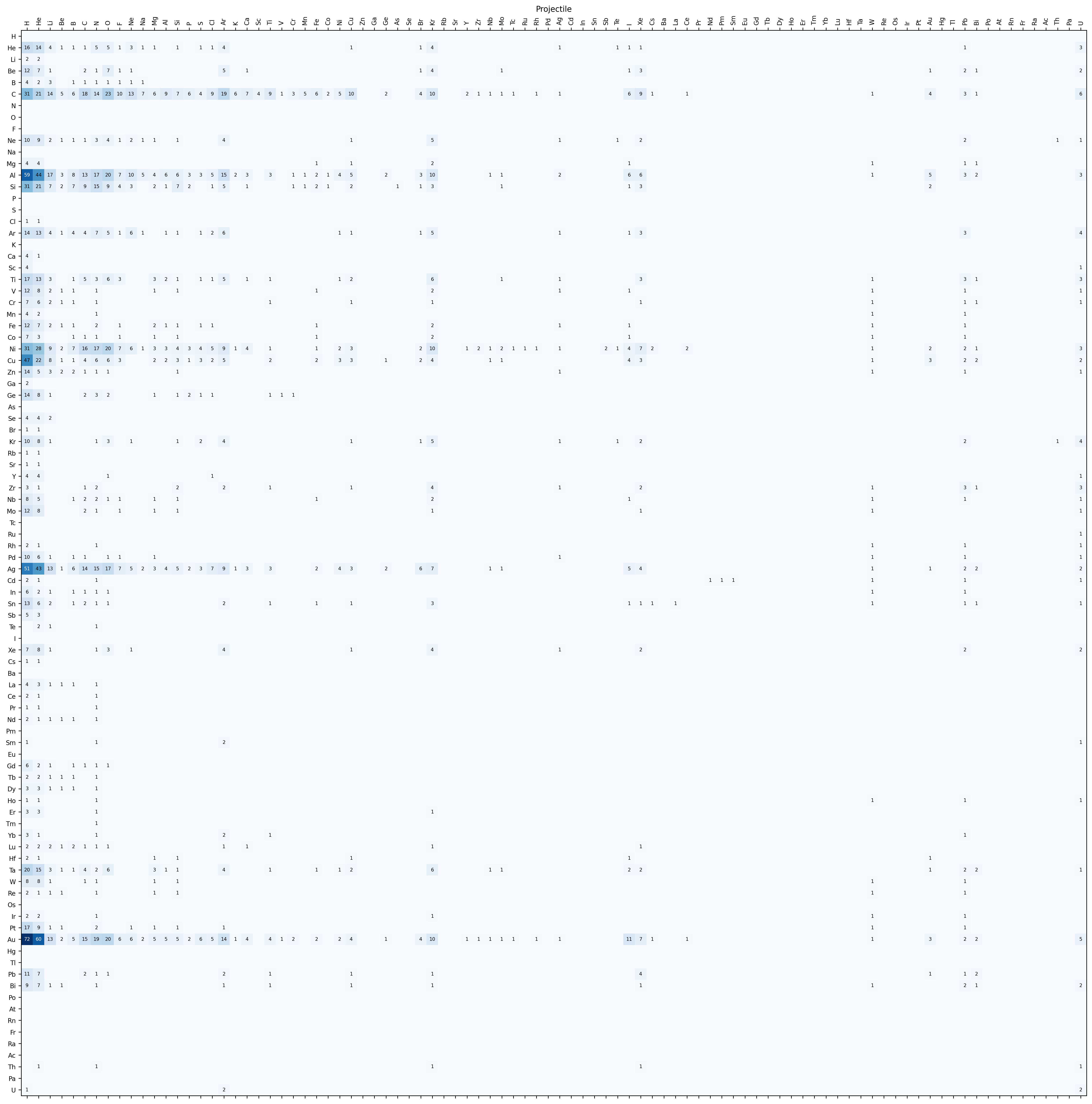}
\caption{Matrix plot with experimental measurement datasets for all the 
collisional systems available in the database.}
\label{fig:matrixplot}
\end{figure}
\begin{figure}[H]\ContinuedFloat
  \centering
  \includegraphics[width=\textwidth]{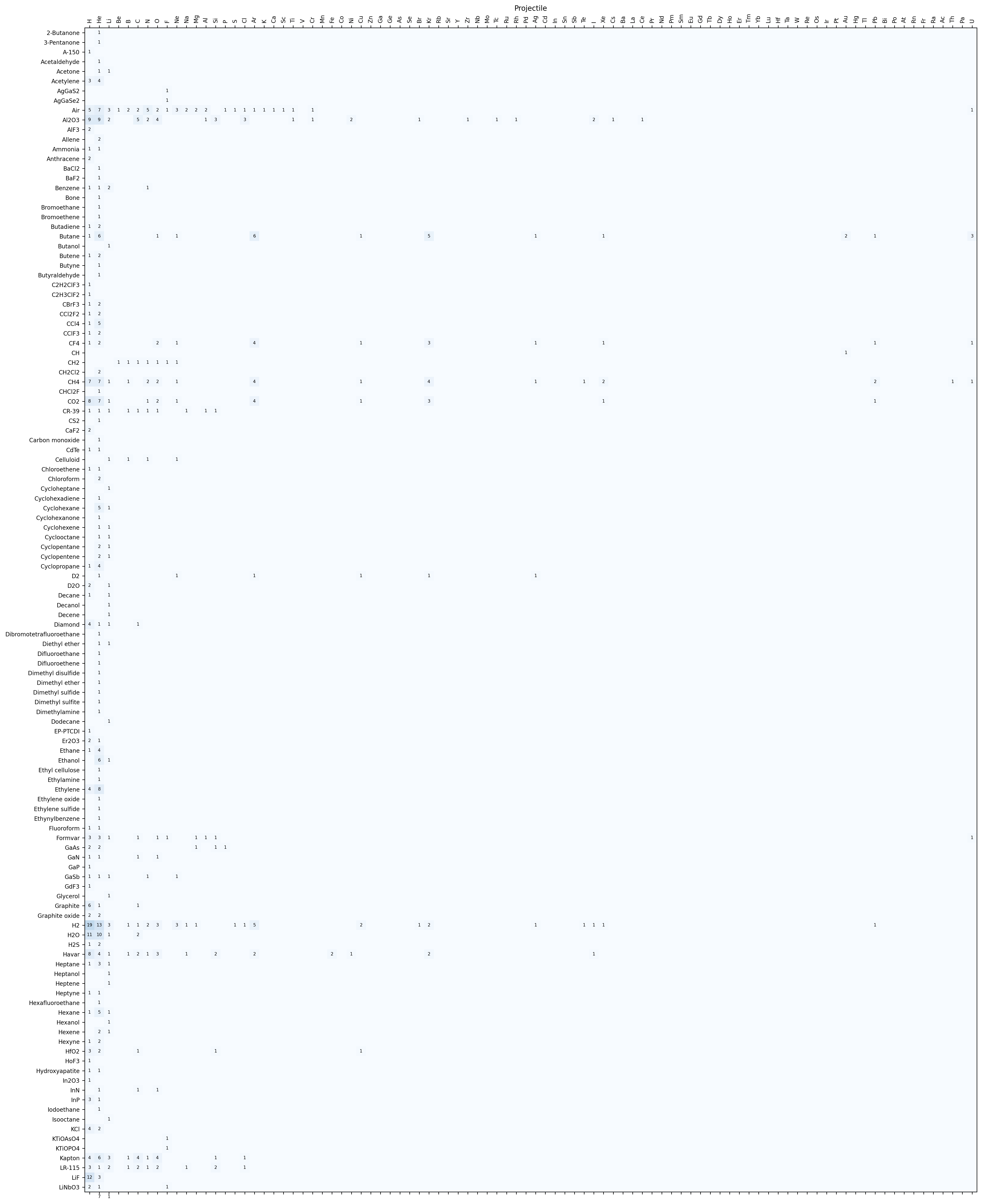}
  \label{fig:comp_part1}
  \caption[]{\textit{(cont.)} Matrix plot with experimental measurement datasets for 
  all the collisional systems available in the database.}
\end{figure}%
\begin{figure}[ht]\ContinuedFloat
  \centering
  \includegraphics[width=\textwidth]{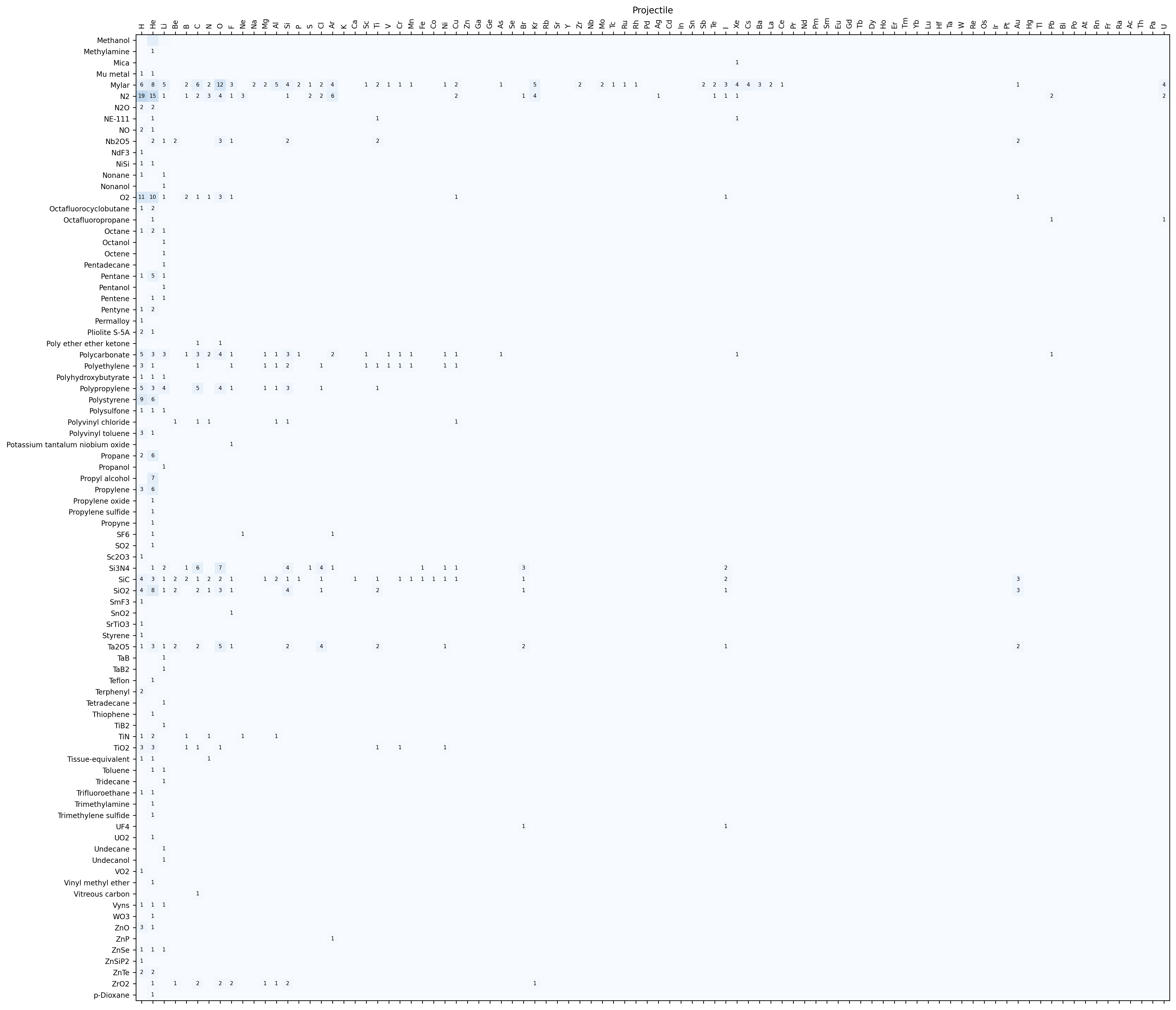}
  \caption[]{\textit{(cont.)} Matrix plot with experimental measurement datasets for 
  all the collisional systems available in the database.}
\end{figure}